# Retrieval of the reflection coefficient in neutron specular reflectometry

S.F. Masoudi and A. Pazirandeh

Physics Department, Tehran University, P.O. Box 1943-19395, Tehran, Iran

**Abstract**
A method is proposed which allows a complete determination of the complex reflection coefficient for any free unknown real potential (i.e., in the case where there is no effective absorption). In this method the unknown layer mounted on top of a known magnetic reference layer in touch with a macroscopic substrate. It lets us to determine the complex reflection coefficient without the effect of the substrate. The method exploits the interference of the spin components of a polarized neutron beam in the presence of a magnetic field. It is based on calculating the polarization of the reflected beam using the elements of transfer matrix and requires the measurement of the polarization in two(one) direction and (two) reflectivity of the reflected beam. A schematic example is presented.

## 1. INTRODUCTION

Specular reflectivity of cold and ultracold neutrons can provide important information about physical and chemical phenomena occurring at surfaces and interfaces of samples in the nanometer range [1-8]. The use of such neutrons reflection to study thin films and interfaces has increased dramatically. In neutron reflection, the interaction of neutrons with the film's atomic constituents is characterized by a continuous scattering length density (SLD) function which is a number density-weighted microscopic average of known isotope specific constant, the scattering lengths. So SLD can determine the coherent elastic scattering behavior of neutrons by the microstructure of a film [9].

The major main of neutron specular reflectometry is to reveal the in-plane average of the SLD depth profile in the direction normal to the surface of materials in thin film geometries [5]. The SLD depth profile of a sample can be directly converted to the chemical profile of the sample [10].

The SLD depth profile determines the specular reflection from a film, so it can be deduced from measurements of the reflectivity R(q), the number of neutrons reflected elastically and specularly as a function of the glancing angle of incidence reflection, relative to the plane of the film or the normal component of the wave vector of the incident neutron.

However, extracting the profile from measured reflectivity has been hampered by the so-called *phase problem* [11-13]. This problem, widely discussed in structure analysis, refers to the fact that in reflection experiment only the square of the complex reflection coefficient r(q), is measured and so like any other scattering technique the phase of reflection is lost [14]. In the absence of the phase, generally least-squares methods [15, 16] are used to extract the SLD profile, but in general more than one SLD may be found to correspond to the same reflectivity [17]. A simple example is the difference between a freestanding SLD profile $\rho(x)$, and its mirror image, $\rho(L-x)$, where L is the thickness of the film and x is depth from the surface. Both produce the same R(q) but different r(q); only the phase of reflection distinguishes one from the other. So given the phase it is possible to solve the one-dimensional inverse scattering problem directly to obtain a unique SLD depth profile [18-21]. So it is first necessary to have a reliable and practical way of determining the reflection coefficient r(q), from reflectivity measurements. Owing to the analytic properties of r(q), there exists a dispersion relation between the phase and the logarithm of the reflectivity which, however does not take into account the contributions of the zeros of the reflection coefficient in the upper half q-plane [22]. Thus the phase problem reduces the problem of determining these zeros [23]. Apart these mathematical considerations several methods for actual determining the phase have been developed like reference layer method [24-35], Lloyd mirage technique [36] and dwell time method [37]. Among these methods the reference layer method (see Ref. 24 as a summary paper) seems the best one because of its application in experiment. The experimental implementation of the reference layer method [25] was first achieved with good success by Majkrzak et al. [26], who also proposed and tested experimentally the related surround method [27,28].

Reference layer methods based on polarization measurements [32-35] are of particular interest, since they also work in the total reflection regime and allow unique reconstruction of surface profiles of magnetic sample. The use of magnetic reference layers has another advantage that the state of layer can be changed without a chemical or structural effect on the sample. In this method r(q) can be determined by measuring reflectivity or polarization of the reflected neutrons. If the reflectivity is measured, it is not important that the reference magnetic layer attaches the front or the back of the unknown film since the SLD or the reveal of SLD can be deduced. In the case we measure the polarization, the method that proposed in Refs. [32-35] is not applicable to determine the reflection coefficient from the back of the an unknown sample because this method use the fact that the wave function for scattering "from the left" by a sample on a substrate of finite thickness can be expressed in terms of the corresponding "left" and "right" scattering functions for a bulk substrate [38]. In this case, H. Leeb et. al. [34], have been shown that the reflection from the right side of the unknown sample can be determined, however here two sets of measurements with different reference layers are necessary and the incident medium must be vacuum.

In this letter we propose a new efficient method for determining r(q) when the unknown film mounted on top of a magnetic reference layer. It has been shown that three measurements of the reflectivity or measuring the polarization in three direction, lead to exact determination of r(q). However we show that to avoid of any problem to find physical solution of r(q), that there is in some methods [e.g. 27,33,37], it is better to measure the polarization of reflected beam in two direction an the reflectivity for up or down polarized neutron. One of the advantages of the method is the possibility of determining the reflection coefficient of the free unknown layer. It is not possible if the reference layer mounted on top of the unknown layer because here the reflection coefficient of the unknown layer plus substrate can be determined [33]. In this paper we assume that the incident medium (fronting) is vacuum. The case of non-vacuum fronting can be investigated very simply.

## 2. THE METHOD

We consider the arrangement of H. Leeb et. al. [34], with a magnetic reference layer (e.g. a Co layer), an unknown sample and a substrate (e.g. a Si wafer) without Cd film because in our method there is any absorption. In Comparison with Refs. [33, 35, 36], we interchange the position of the sample and the reference layer, as shown in Fig. 1 (i.e. in our arrangement the magnetic reference layer is in touch with the substrate). The reference layer is magnetized in a direction parallel to the reflecting surface (+z direction) which is taken as the direction of spin quantization. It is assumed that the external magnetic field vanishes in the unknown sample. However it is obvious that for non-magnetic unknown layers this assumption is not necessary.

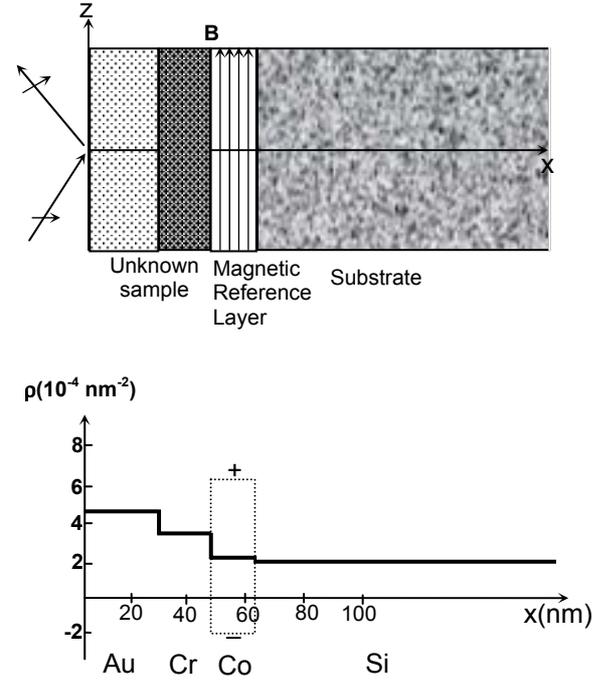

Fig. 1. Experimental arrangement for measuring the complex reflection coefficient. Top: arrangement of the layers. Bottom: the scattering length density depth profile. The dotted lines represent the effective scattering length density experienced by neutron beams polaraized parallel and antiparallel to the magnetic field **B**.

For incident beam polarized in the direction of the normal to the reflection surface (+x-direction), the polarization of the reflected beam can be expressed as follow:

$$P_z = \frac{R_+ - R_-}{R_+ + R_-} \qquad (1)$$

$$P_x + iP_y = \frac{2r_+^* r_-}{R_+ + R_-} \qquad (2)$$

where $r_\pm(q)$ are the reflection coefficients for neutron beam polarized parallel (+) or antiparallel (-) to the magnetic field (or plus and minus magnetization of reference layer) respectively and $R_\pm = |r_\pm|^2$.

$r_\pm(q)$ can be derived from the transfer matrix method of solving a one-dimensional Schrodinger equation

$$\partial_x^2 \Psi(q,x) + 4\pi\rho(x)\Psi(q,x) = q^2\Psi(q,x) \qquad (3)$$

The 2×2 unimodular transfer matrix has elements A(q), B(q), C(q) and D(q) that are completely determined by knowing the SLD depth profile of sample, $\rho(x)$.

In our sample SLD profile is separated into two distinct region representing known magnetic and unknown parts. So the total transfer matrix can be expressed as a product of corresponding transfer matrices

$$\begin{pmatrix} A & B \\ C & D \end{pmatrix} = \begin{pmatrix} w_\pm & x_\pm \\ y_\pm & z_\pm \end{pmatrix} \begin{pmatrix} a & b \\ c & d \end{pmatrix} \quad (4)$$

where the matrix (a,…,d) describes the contribution from the unknown part of $\rho$ and ($w_\pm,…,z_\pm$) gives the known part, i.e. magnetic reference layer. (+) and (-) denote the plus and minus magnetization of the reference layer respect to neutron beam polarized parallel and anti parallel to the magnetic field. It is assumed that absorption in all film materials as well as in the substrate is negligible. An important consequence of this is that the transfer matrix elements are real-valued at all q.

In transfer matrix method it is easier to treat with three functions

$$\alpha^h = hA^2 + h^{-1}C^2 \quad (5a)$$
$$\beta^h = hB^2 + h^{-1}D^2 \quad (5b)$$
$$\gamma^h = hAB + h^{-1}CD \quad (5c)$$

where "h" as subscript denotes that sample is mounted on top of a substrate having constant SLD $\rho_s$ and

$$h = (1 - 4\pi\rho_s/q^2)^{1/2} \quad (6)$$

Using relation 4, these parameters can be expressed as

$$\alpha_\pm = \alpha_\pm^h a^2 + \beta_\pm^h c^2 + 2\gamma_\pm^h ac \quad (7a)$$
$$\beta_\pm = \alpha_\pm^h b^2 + \beta_\pm^h d^2 + 2\gamma_\pm^h bd \quad (7b)$$
$$\gamma_\pm = \alpha_\pm^h ab + \beta_\pm^h cd + 2\gamma_\pm^h (ad+bc) \quad (7c)$$

where $\alpha_\pm$, $\beta_\pm$ and $\gamma_\pm$ are denoted for whole arrangement containing known film, magnetic reference layer and substrate, and $\alpha_\pm^h$, $\beta_\pm^h$ and $\gamma_\pm^h$ are respect to the magnetic reference layer mounted on top of the substrate. The reflectivity, $R_\pm$, can be related to these functions in terms of a new quantity $\Sigma_\pm(q)$,

$$\Sigma_\pm = 2\frac{1+R_\pm}{1-R_\pm} = \alpha_\pm + \beta_\pm \quad (8)$$

(This relation is different respect to the same quantity that introduced by Majkrzak et. al. [24], except in h quantity)

Using Eqs. (7a) and (7b), we have

$$\Sigma_\pm(q) = \beta_\pm^h \tilde{\alpha}_u + \alpha_\pm^h \tilde{\beta}_u + 2\gamma_\pm^h \tilde{\gamma}_u \quad (9)$$

where the tilde denotes a reversed unknown film, that is the interchange of the diagonal elements of the corresponding transfer matrix (A,…,D→d,b,c,a).

The complex reflection coefficient, $r_\pm(q)$ can expressed using these parameters by:

$$r_\pm(q) = \frac{\beta_\pm - \alpha_\pm - 2i\gamma_\pm}{\Sigma_\pm + 2} \quad (10)$$

For free reversed unknown film, i.e. without reference layer and substrate, this equation reduces to

$$r(q) = \frac{\tilde{\beta}_u - \tilde{\alpha}_u - 2i\tilde{\gamma}_u}{\tilde{\beta}_u + \tilde{\alpha}_u + 2} \quad (11)$$

This indicates that knowing $\tilde{\alpha}$, $\tilde{\beta}$ and $\tilde{\gamma}$ the complex reflection coefficient of the reversed film can be deduced.

To show the ability of finding these parameters we first express the polarization of the reflected beam by these parameters. Using Eqs. (1), (2), (7) and (10), it seems that Eqs. 1 and 2 are very complicated but by some calculation, one can find that

$$P_x = 1 + \frac{\zeta}{\Sigma_+ \Sigma_- - 4} \quad (12)$$

$$P_y = \frac{2}{\Sigma_+ \Sigma_- - 4}\{(\gamma_+^h \beta_-^h - \beta_+^h \gamma_-^h)\tilde{\alpha}_u + (\alpha_+^h \gamma_-^h - \gamma_+^h \alpha_-^h)\tilde{\beta}_u$$
$$+ (\alpha_+^h \beta_-^h - \beta_+^h \alpha_-^h)\tilde{\gamma}_u\} \quad (13)$$

$$P_z = 2\frac{\Sigma_+ - \Sigma_-}{\Sigma_+ \Sigma_- - 4} \quad (14)$$

where

$$\zeta = 4(1 + \gamma_+^h \gamma_-^h) - 2(\alpha_+^h \beta_-^h + \beta_+^h \alpha_-^h) \quad (15)$$

The ability of our method is based on above equations since in Eqs. 12 and 14, $P_x$ and $P_z$ are associated with three unknown parameters by $\Sigma_\pm$, and in Eq. 13 $P_z$, without considering $\Sigma_\pm$, is associated by linear combination of the unknown parameters. There aren't so conditions when the magnetic reference layer is mounted on top of the unknown layer.

As an example, we consider the case that the polarization of the reflected beam is measured. By using Eqs. (12) and (14) we can deduce $\Sigma_+$ and $\Sigma_-$ as follow:

$$\Sigma_\pm^2 \mp \frac{\zeta P_z}{2(P_x - 1)}\Sigma_\pm - (4 + \frac{\zeta}{P_x - 1}) = 0 \quad (16)$$

So Eqs. (9) and (13) can be used to find three unknown parameters. Since Eq. 16 has two roots the physical solution must be selected. However it is clear that physical solution must satisfy $\Sigma_\pm > 2$. This problem is seen in some another methods [24,33,34]. The difference between our method and the one proposed by H. leeb et. al. [34] is that in their method the reflection coefficient from the right by unknown sample is determined whereas in our method the reflection coefficient from the left of free unknown film can be determined.

To avoid of having two solutions for $r_\pm(q)$, we suggest to measure combination of the reflectivity and polarization; ($\Sigma_+$, $\Sigma_-$ and $P_y$), (one of $\Sigma_+$ or $\Sigma_-$, $P_y$ and $P_z$) or (one of $\Sigma_+$ or $\Sigma_-$, $P_x$ and $P_y$). Existence of $P_y$ in all cases shows that measuring the polarization in y-direction (parallel to the sample surface and perpendicular on external magnetic filed) is necessary.

For example if we measure the reflectivity for an incident polarized neutron beam antiparallel to the magnetic field and $P_x$ and $P_y$ for incident polarized neutron beam normal to the surface of the sample (+x-

direction) three unknown parameters can be obtained as follow:

$$\begin{pmatrix} \tilde{\alpha}_u \\ \tilde{\beta}_u \\ \tilde{\gamma}_u \end{pmatrix} = M^{-1} \begin{pmatrix} \Sigma_- \\ \dfrac{1}{\Sigma_-}(4+\dfrac{\zeta}{P_x-1}) \\ \zeta P_y / 2(P_x-1) \end{pmatrix} \quad (17)$$

where

$$M = \begin{pmatrix} \beta^h_- & \alpha^h_- & 2\gamma^h_- \\ \beta^h_+ & \alpha^h_+ & 2\gamma^h_+ \\ \gamma^h_+\beta^h_- - \beta^h_+\gamma^h_- & \alpha^h_+\gamma^h_- - \gamma^h_+\alpha^h_- & \alpha^h_+\beta^h_- - \beta^h_+\alpha^h_- \end{pmatrix} \quad (18)$$

The case of non vacuum baking can be investigated easily from above equations whereas the method proposed by H. Leeb et. al. [34] to determine r(q) from right side of the unknown sample can not be used for non-vacuum fronting. In this case, all the known parameters $\alpha^h_\pm$, $\beta^h_\pm$ and $\gamma^h_\pm$ change to $\alpha^{fh}_\pm$, $\beta^{fh}_\pm$ and $\gamma^{fh}_\pm$. The "fh" subscript indicates that the reference layer is between the fronting and substrate and also $\alpha^{fh}_\pm = f\alpha^h_\pm$, $\beta^{fh}_\pm = f^{-1}\beta^h_\pm$ and $\gamma^{fh}_\pm = \gamma^h_\pm$ where f is defined like Eq. 5 with the SLD of fronting media.

**Example**

As a realistic example to test the method by simulation, we consider the setup of Fig. 1. The sample is a 30nm thick gold on a 20nm Cr having constant SLD values of $4.46\times10^{-4}$ nm$^{-2}$ and $3.03\times10^{-4}$ nm$^{-2}$ respectively. The magnetic reference layer is a 15nm thick Co layer having SLD values of $6.44\times10^{-4}$nm$^{-2}$, $-1.98\times10^{-4}$nm$^{-2}$ and $2.23\times10^{-4}$nm$^{-2}$ respect to plus, minus and non magnetization. The magnetization of this layer will also generate a magnetic induction outside the ferromagnetic film, which we assume, however, to be small enough not to affect the neutron beam. For simplicity, we set it equal to zero. Absorption and roughness of interfaces are neglected.

The reflectivity $R_\pm$ for incident beam to be fully polarized in the $\pm z$-direction, has been calculated in Fig. 2a. Figs. 2b-2d show the q dependence of the polarization of reflected beam, $P_x$, $P_y$ and $P_z$ for incident beam to be fully polarized in the +x-direction. In all calculation the simulated data start at the critical q of the substrate.

Figs. 3a and 3b show the real and imaginary part of the reflection coefficient for mirror image of the unknown sample (i.e. Cr layer is mounted on top of the gold layer) derived from the data of Fig. 2a (blue), 2b and 2c and using Eq. 17.

At the below of the critical q for total external reflection by the substrate the fact that Rer(q)→-1 and Imr(q)→0 makes it possible to reliably interpolate in this region [26].

Once r(q) has been measured, it can be inverted for the desired ρ(L-x) by solving the Gel'fand-Levitan-Marchenko integral equations. However, since we measured r(q) from the left side of the sample it is not need to change the arrangement of the evaluated potential from x to (-x) that is need in the method of H.Leeb et. al. [34].

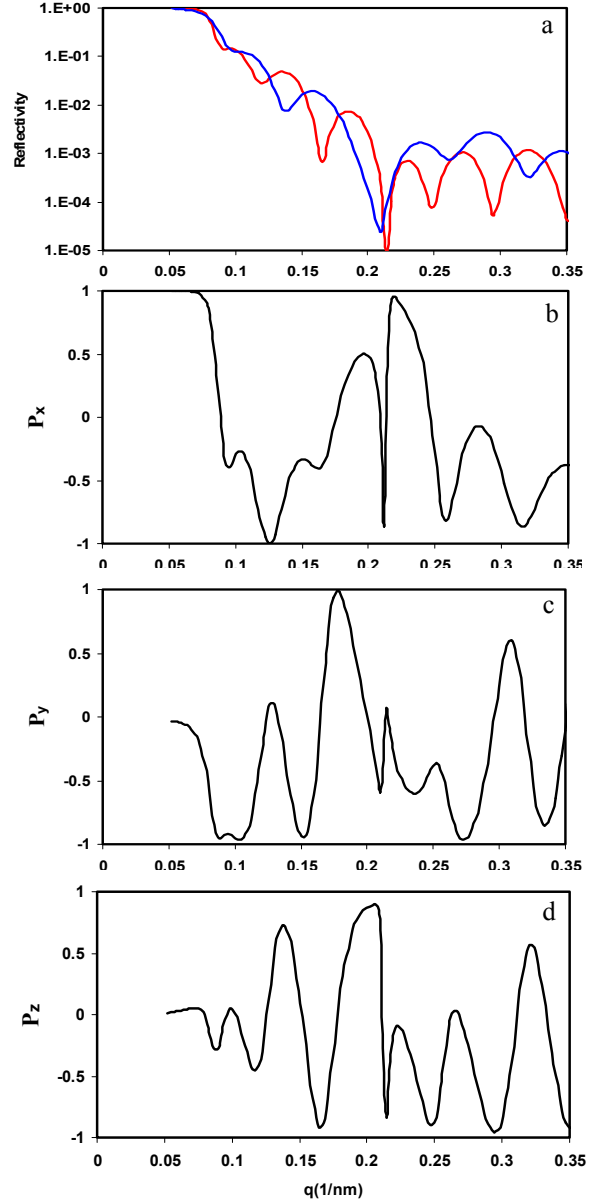

Fig. 2. Simulated reflectivity and polarization data for the arrangement of Fig. 1. The reflectivity for up(red) and down(blue) polarized incident beam fully in ±z direction (a), and the polarization components $P_x$ (b), $P_y$ (c) and $P_z$ (d) of the reflected beam are shown. For polarization, the incident beam is assumed to be fully polarized in +x direction. The simulate data start at the critical q of the substrate.

Since the method is strongly related to the reference layer method, its stability with respect to experimental uncertainties, e.g., roughness of interfaces and measurement errors, is similar to that of the method of H. Leeb et. Al [33].

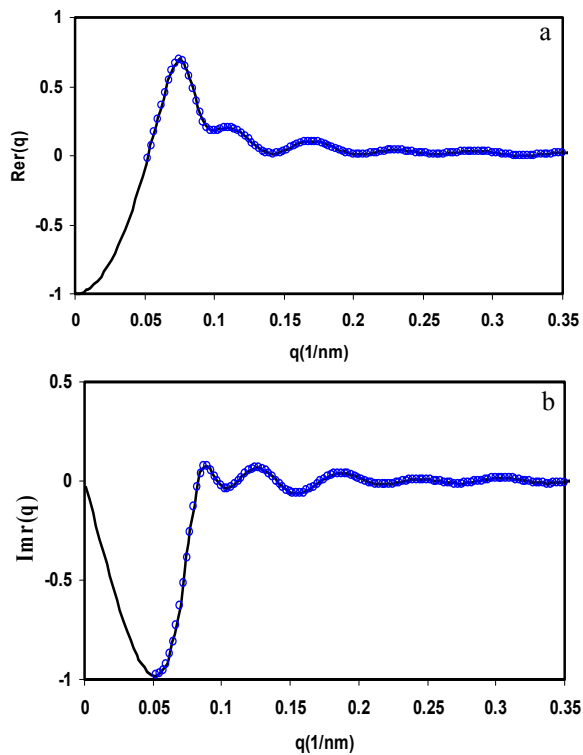

Fig. 3. Real part (a) and Imaginary part (b) of the mirror image of the unknown film of Fig. 1 (i.e. the Cr film is mounted on top the gold film). Solid line: computed directly from Eq. 11. Blue circles: recovered from data of Fig. 2a (blue ) 2b and 2c.

## Conclusions

Without knowledge of the phase of reflection coefficient or without sufficient additional physical information, a single set of reflectivity data for a nonsymmetric sample film alone can only reveal whether a particular SLD profile in consistent with the data, but not that is the only possible result. The phase of the reflection coefficient can be used as an additional information to guarantee the uniqueness of a SLD profile. So knowing the reflection coefficient $r(q)$ is necessary.

We have proposed a method to determine $r(q)$ using a magnetic reference layer in touch with a macroscopic substrate. The unknown layer is mounted on top of the reference layer. Compared to the standard reference layer method, which is based on the measurement of reflectivities or polarization (both of these methods are applicable with our method) we have shown that it is better two measuring the polarization in two direction $(P_x, P_y)$ or $(P_y, P_z)$ with incident beam polarized normal to the sample surface (+x-direction) and the reflectivity $R_+$ or $R_-$ with incident beam polarized parallel (+) or anti-parallel (-) to the external magnetic field. The case of measuring $P_y$, $R_+$ and $R_-$ is applicable too. However the formula found in the letter can be used to determining the complex reflection coefficient in the case that three reflectivity, or the polarization of the reflected beam is measured.

**References**

[1] J.B. Hayter, R.R. Highfield, B.J. Pullman, R.K. Thomas, A.I. McMullen and J. Penfold, J. Chem. Soc. Faraday Trans. I 97 (1981) 1437.
[2] G. Felcher, Proc. SPIE 983, 2 (1988).
[3] T.P. Russel, Matter. Sci. Rep. 5, 171 (1990).
[4] J. Penfold and R.K. Thomas, J. Phys. Condens. Matter 2, 1369 (1990).
[5] Workshop on Methods of Analysis and Interpretation of Neutron Reflectivity Data, edited by G. Felcher and T.P. Russel (Physica B 173, 1 (1991)).
[6] H.J. Lauter and V.V. Pasyuk, Proceeding of the International Conference on Surface X-ray and Neutron scattering [Physica B 198, 1 (1994)].
[7] G.P. Felcher, Neutron News 5,18 (1994).
[8] C.F. Majkrzak, in Neutron Scattering im material Science II, edited by D.A. Neumann, T.P. Russell and B.J. wuensch, MRS Symposia Proceeding No. 376 (Materials research Society, Pittsburgh, 1995).
[9] V.F. Sears, Neutron Optics, (Oxford University Press), 1989.
[10] V.F. Sears, in: Methods of experimental physics, Vol. 23, Neutron Scattering, eds. K. Skold and D.L. Price (Academic Press, Orlando, 1986) 521.
[11] G.Reiss and R. Lipperheide, Phys. Rev. B 53, 8157 (1996).
[12] H. Fiedeldey, R. Lipperheide, H. Leeb and S.A. Sofianos, Phys. Lett. A 179 (1992) 347.
[13] P.S. Pershan, Phys. Rev. E 50 (1994) 2369.
[14] H.A. Hauptmann, Rep. Prog. Phys. 54, 1427 (1991).
[15] V.O. de Haan and G.G. Drijkoningen, Physica B 198, 24 (1994).
[16] X.L. Zhou and S.H. Chen, Phys. Rev. E 47, 121 (1993).
[17] R.E. Burge, M.A. Fiddy, A.H. Greenaway, and G.Ross, Proc. R. Sco. London Ser. A 350, 191 (1976).
[18] B.N. Zakhariev and A.A. Suzko, Direct and Inverse Problems (Springer-Verlag, Heidelberg, 1990).
[19] T.M. Reports, Physica B 173, 143 (1991).
[20] P.E. sacks, wave Motion 18, 21 (1993).
[21] K. Chadan and P.C. Sabatier, Inverse Problem in Quantum Scattering Theory, 2$^{nd}$ ed. (Springer, New York, 1989).
[22] M.V. Klibanov, P.E. Sacks, J. Math. Phys. 33 (1992) 3813; J.Comp. Phys. 112 (1994) 273.
[23] R. Lipperheide, G. Reiss, H. Leeb and S.A. Sofianos, Physica B 221 (1996) 514.
[24] C.F. Majkrzak, N.F. Berk, and U.A. Perez-Salas, Langmuir 2003, 19, 7796-7810 (In this paper a comprehensive summary of the experimental and theoretical methodologies of phase-sensitive neutron reflectometry is presented and illustrated by a current application).
[25] C.F. Majkrzak, N.F. Berk, Phys. Rev. B 52, 10 827 (1995).
[26] C.F. Majkrzak, N.F. Berk, J. Dura, S. Atija, A. Karim, J. Pedulla, and R.D. Deslattes, Physica B 248, 338 (1998).
[27] C.F. Majkrzak, N.F. Berk, Physica B 221, 520 (1996); Phys. Rev. B 58, 15 416 (1998).
[28] C.F. Majkrzak, N.F. Berk, V. Silin, and C.W. Meuse, Physica B 283, 248 (2000).
[29]V.O. de Haan, A.A. van Well, S. Adenwalla, and G.P. Felcher, Phys. Rev. B. 52, 10 831 (1995).
[30] H. Leeb, R. Lipperheide, G. Reiss, J. Phys. Soc. Jpn., Suppl. A 65 (1996) 138.
[31] R. Lipperheide, M. Weber, H. Leeb, Physica B 283 (2000) 242-247.

[32] R. Lipperheide, J. Kasper, H. Leeb and G. Reiss, Physica B 234-236 (1997) 1117-1119.
[33] H. Leeb, J. Kasper, and R. Lipperheide, Phys. Lett. A 239 (1998) 147-152.
[34] H. Leeb, H. Groz, J. Kasper and R. Lipperheide, Phys. Rev. B. 63, (2001) 045414.
[35] J. Kasper, H. Leeb, and R. Lipperheide, Phys. Rev. Lett., 80 (1998) 2614.
[36] V.P. Gudkov, G.I. Opat, A.G. Klein, J. Phys.: Condensed Matter 5 (1993) 9013.
[37] H. Fiedeldey, R. Lipperheide, H. Leeb and S.A. Sofianos, Phys. Lett. A 170 (1992) 347-351.
[38] K. Chadan and P.C. Sabatier, Inverse Problems in quantum scattering theory, 2$^{nd}$ Ed. (Springer, Berlin, 1989).